\documentclass[prl,aps,showpacs,twocolumn]{revtex4}
\usepackage{epsfig,color}
\usepackage{bm} 

\renewcommand{\it}[1]{\textit{#1}}
\newcommand{\Onecol} {\begin{widetext} \onecolumngrid} 
\newcommand{\Twocol} {\end{widetext} \twocolumngrid} 

\newcommand{\be}{\begin{equation}}
\newcommand{\ba}{\begin{array}}
\newcommand{\bea}{\begin{eqnarray}}
\newcommand{\bfi}{\begin{figure}}
\newcommand{\ee}{\end{equation}}
\newcommand{\ea}{\end{array}}
\newcommand{\eea}{\end{eqnarray}}
\newcommand{\efi}{\end{figure}}

\begin{document} 
\title{Magnetic reversals in a simple model of MHD}
\author{Roberto Benzi${}^{(1)}$, Jean-Fran\c{c}ois Pinton${}^{(2)}$}
\affiliation{(1) Dip. di Fisica and INFN, Universit\`a ``Tor Vergata"\\
Via della Ricerca Scientifica 1, I-00133 Roma (Italy)\\ 
(2) Laboratoire de Physique, CNRS \& \'Ecole Normale Sup\'erieure de Lyon, \\ 
46 all\'ee d'Italie, F69007 Lyon (France)}

\begin{abstract} 
We study a simple magnetohydrodynamical approach in which hydrodynamics and MHD turbulence are coupled in a shell model, with given dynamo constrains in the large scales. We consider the case of a low Prandtl number fluid for which the inertial range of the velocity field is much wider than that of the magnetic field. Random reversals of the magnetic field are observed and it shown that the magnetic field has a non trivial evolution -- linked to the nature of the hydrodynamics turbulence.
\vskip 0.2cm 
\end{abstract} 


\maketitle 

Observations show that natural dynamos are intrinsically dynamical. Complex magnetic field evolutions have been reported for many systems, including the Sun and the Earth~\cite{RobertsBook}. Formally, the coupled set of momentum and induction equations are invariant under the transform: $(\mathbf{u}, \mathbf{B}) \rightarrow (\mathbf{u}, -\mathbf{B})$ so that states with opposite polarities can be generated from the same velocity field ($\mathbf{u}$ and $\mathbf{B}$ are respectively the velocity and magnetic fields). In the case of the geodynamo, polarity switches are called reversals~\cite{RobertsBook} and occur at very irregular time intervals~\cite{Merill}. Such reversals have been observed recently in laboratory experiments using liquid metals, in arrangements where the dynamo cycle is either favored artificially~\cite{BVK} or stems entirely from the fluid motions~\cite{VKSP1,VKSP2}. In these laboratory experiments, as also presumably in the Earth core, the ratio of the magnetic diffusivity to the viscosity of the fluid (magnetic Prandtl number $P_M$) is quite small. As a result, the kinetic Reynolds number $R_V$ of the flow is very high because its magnetic Reynolds number $R_M=R_V P_M$ needs to be large enough so that the stretching of magnetic fields lines balances the Joule dissipation. Hence, the dynamo process develops over a turbulent background and in this context, it is often considered as a problem of `bifurcation in the presence of noise'. For the dynamo instability, the effect of noise enters both additively and multiplicatively, a situation for which a complete theory is not currently available. Some specific features  have been ascribed to its onset (e.g. bifurcation via an {\it on-off}  scenario~\cite{Ott}) and to its dynamics~\cite{Hoyng}. Turbulence also implies that processes occur over an extended range of scales; however, in a low magnetic Prandtl number fluid the hydrodynamic range of scales is much wider than the magnetic one. In laboratory experiments, the induction processes that participate in the dynamo cycle involve the action of large scale velocity gradients~\cite{Riga,Karlsruhe,VKSP1}, with also possible contributions of velocity fluctuations at small scales~\cite{Volk2006,Forest2007,Denisov2008}. 

Building upon the above observations, we propose here a simple model which incorporates hydromagnetic turbulent fluctuations (as opposed to `noise') in a dynamo instability. The models stems from the approach introduced in\cite{benzi05} for the hydrodynamic studies. Magnetic field reversals are observed above onset and we detail their characteristics.

\begin{figure}[h]
	\begin{center}
		\includegraphics[width=0.40\textwidth]{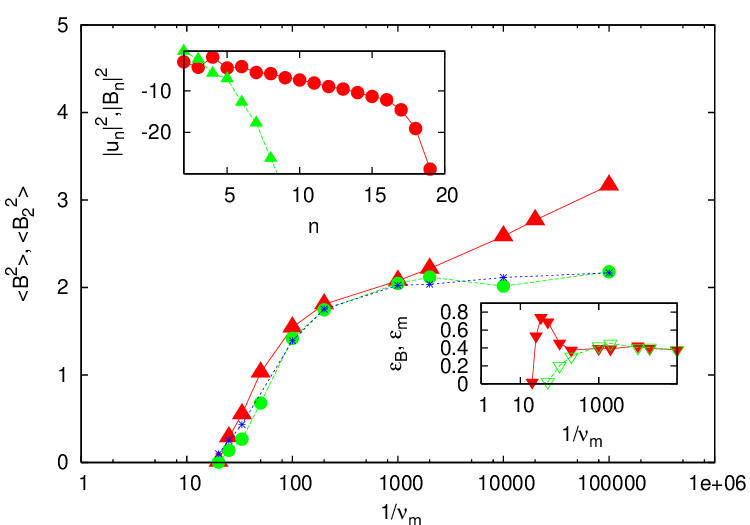}
	\end{center}
	\caption{
	Main figure: the behavior of  $\langle |B_2|^2 \rangle$ (red triangles) and $\langle |B_2|^2\rangle $ (green circles) as a function of $\nu_m$ for fixed value of $\nu=10^{-7}$. The blue line with dot corresponds
	to the solutions of (\ref{A}) . Upper insert: energy spectra for $B_n$ (green triangles) and $u_n$ (red circles)        corresponding to the case $\nu_m= 0.001$.  
	Lower insert: the amount of magnetic dissipation (red triangles) $\epsilon_B =\langle \Sigma_n \nu_m k_n^2 |B_n|^2 \rangle$ and the dissipation due to the large scale term $ \epsilon_m = a_m \langle |B_2|^2(B_{2r}^2-B_{2i}^2) \rangle$ (green triangles).
         }
	\label{fig1}
\end{figure}

We  consider an  `energy cascade' model {\it i.e.} a shell model aimed at reproducing few of the relevant characteristic features of the  statistical properties of the Navier-Stokes equations \cite{luca}. In a shell models, the basic variables describing the `velocity field' at scale  $r_n = 2^{-n} r_0 \equiv k_n^{-1}$, is a complex number $u_n$ satisfying a suitable set of non linear equations. There are many version of shell models which have been introduced in literature. Here we choose the one referred to as {\em Sabra} shell model. Let us remark that the statistical properties of intermittent fluctuations, computed either using shell variables or the instantaneous rate of energy dissipation, are in close {\em qualitative} and {\em quantitative} agreement with those measured in laboratory experiments, for homogeneous and isotropic turbulence \cite{luca}. MHD shell model -- introduced in~\cite{Frick1998} -- allow a description of turbulence at low magnetic Prandtl number since the steps of both cascades can be freely adjusted~\cite{Stepanov2006, Stepanov2007}. Although geometrical features are lost, this is a clear advantage over 3D simulations~\cite{Ponty2004,Baerenzung2008}.
We consider here a formulation extended from the Sabra hydrodynamic shell model: 
\begin{eqnarray}
\frac{du_n}{dt}  & = & \frac{i}{3} (\Phi_n(u,u) - \Phi_n(B,B)) - \nu k_n^2 u_n + f_n\ , \label{1}\\
\frac{dB_n}{dt} & = & \frac{i}{3} (\Phi_n(u,B) - \Phi_n(B,u)) - \nu_m k_n^2 B_n   \ , \label{2}
\end{eqnarray}
where \begin{eqnarray}
&& \Phi_n(u,w) = k_{n+1}[(1+\delta) u_{n+2}w_{n+1}^*+(2-\delta)u_{n+1}^*w_{n+2}]\nonumber\\
&&+ k_{n}[(1-2\delta)u_{n-1}^*w_{n+1}-(1+\delta)u_{n+1}w_{n-1}^*]\nonumber\\
&&+ k_{n-1}[(2-\delta)u_{n-1}w_{n-2}+(1-2\delta)u_{n-2}w_{n-1}] \ ,
\end{eqnarray}
for which  following \cite{benzi05} we chose $\delta= -0.4$. For this value of $\delta$, the {\it Sabra} model is known to show statistical properties (i.e. anomalous scaling) close to the ones observed in  homogenous and isotropic turbulence.  The model, without forcing and dissipation, conserve the kinetic energy $\Sigma_n |u_n|^2$, the magnetic energy $\Sigma_n |B_n|^2$ and the helicity $Re(\Sigma_n u_n B_n^*)$. In the same limit, the model has a $U(1)$ symmetry corresponding to a phase change $\exp(i\theta)$ in both complex variables $u_n$ and $B_n$. The quantity $\Phi_n(v,w)$ is the shell model version of the transport term $\vec{v}\nabla\vec{w}$. The forcing term $f_n$ is given by $f_n \equiv \delta_{1n}f_0/u_1^*$, {\it i.e.} we force with a constant power injection in the large scale.    We want to introduce in eq. (\ref{2}) an extra (large scale) term aimed at producing two statistically stationary equilibrium solutions for the magnetic field. For this purpose, we add to the r.h.s. of (\ref{2}) an extra term $M_2(B_2)$,
namely for $n=2$ eq.(\ref{2}) becomes:
\begin{equation}
\label{b2}
\frac{dB_2}{dt} = F_2(u,B) - M_2(B_2)-\nu_m k_2^2 B_2
\end{equation}
where $F_2(u,B)$ is a short hand notation for $i/3 (\Phi_2(u,B)-\Phi_2(B,u))$. The term $M_2(B_2)$ is chosen with two requirements: 1) it must break the   $U(1)$ symmetry; 2) it must introduce a large scale dissipation needed to equilibrate the large scale magnetic field production. There are many possible ways to satisfy these two requirements. Here we simply choose $M_2(B_2)=a_m B_2^3$. We argue, see the discussion at the end of this letter, that the two requirements are a necessary condition to observe large scale equilibration.  From a physical point of view, symmetry breaking also occurs in real dynamos since the magnetic field is directed in one preferential direction which changes sign during a reversal. Thus symmetry breaking is a generic feature which {we introduce} in our model by prescribing some large scale geometrical constrain. On the other hand, large scale dissipation must be responsible of the equilibration mechanism of the large scale field. The choice of a non linear equilibration is made here to highlight the the existence of a non linear center manifold for the large scale dynamics. In other words, eq.(4) with $M_2(B_2) = a_m B_2^3$ is supposed to describe the `normal form' dynamics of the large scale magnetic field. Note, that our assumption on $M_2$ does not necessarily imply  a time scale separation between the characteristic time scale of $B_2$ and the magnetic turbulent field. Finally,  since the  system has  an inverse cascade of helicity, we set $B_1=0$ as boundary conditions.

The free parameters of the model are the power input $f_0$, the magnetic viscosity $\nu_m$ and the saturation parameters $a_m$.  Actually, the parameter $f_0$ could be eliminated by a suitable rescaling of the velocity field. We shall keep it fixed  to $f_0 = 1-i$.  In figure (\ref{fig1}) we show the amplitude of $\langle |B_2|\rangle $  and the magnetic energy $E_B \equiv \langle =\Sigma_n |B_n|^2\rangle$ as a function of $\nu_m$ for $\nu=1e-7$, where the symbol $\langle .. \rangle$ stands for time average . For very large $\nu_m$, the magnetic field does not grow. Then, for $\nu_m $ greater than some critical value, $\langle B_2 \rangle$ as well as $E_B$ increases for decreasing $\nu_m$. Eventually, $\langle |B_2| \rangle $ saturates at a given value while $E_B$ still increases, showing that for $\nu_m$ small enough a fully developed spectrum of $B_n$ is achieved. This {type of} behavior  is in agreement with previous {studies of}  Taylor-Green flows~\cite{Ponty05,Laval06}, $s_2t_2$ flows in a sphere~\cite{Bayliss07} or MHD shell models~\cite{Frick06}. In the top insert of the same figure we show the magnetic and energy spectrum for $\nu_m = 10^{-3}$. Finally, in the lower insert we plot the magnetic dissipation $\epsilon_B =\nu_m \Sigma_n k_n^2 \langle |B_n|^2 \rangle$ and the large scale dissipation due to $M_n$. Note, that at the dynamo threshold, we observe a sudden bump in the magnetic dissipation which decreases for decreasing $\nu_n$. At relatively small $\nu_m$, the magnetic dissipation becomes constant and quite close to the large scale dissipation. 

We can reasonably predict the behaviour of $\langle |B_2|^2 \rangle$ as function of $\nu_m$ by the following argument. The onset of dynamo implies that there exists a net flux of energy from the velocity field to the magnetic field.  At the largest scale, the magnetic field $B_2$ is forced by the velocity field due to the terms $F_2(u,B)$, see eq.(4). The quantity $A \equiv {\cal R}[F_2(u,B)B_2^*]$ is the energy pumping due to the velocity field which is independent on $B_2$ and $a_m$. Thus, from eq.(4) we can obtain:
\begin{equation}
\label{A}
\frac{1}{2} \frac{d|B_2|^2}{dt} = A - a_m |B_2|^2(B_{2r}^2-B_{2i}^2)- \nu_m k_2^2 |B_2|^2
\end{equation}
where $B_{2r}$ and $B_{2i}$ are the real and imaginary part of $B_2$. For large $\nu_m$, the amplitude of $B_2$ is small and the symmetry breaking term proportional to $a_m$ is negligible. Under this condition, and with the boundary condition constrains, we expect from (\ref{A}) or (\ref{b2}) that the behavior of $B_2$ is periodic, as it has been observed in the numerical simulations. On the other hand for relatively small $\nu_m$, the non linear equilibration breaks the U(1) symmetry and $B_{2i}$ becomes rather small and statistically stationary solutions can be observed with $B_{2r}^2 = \sqrt{A/a_m}$. Computing $A$ from the numerical simulations, we can use (\ref{A}) to predict how $\langle |B_2|^2 \rangle$ depends on $\nu_m$. The results is shown in figure~\ref{fig1} by the blue line with rather good agreement.
\begin{figure}[t]
\centerline{\includegraphics[width=8cm]{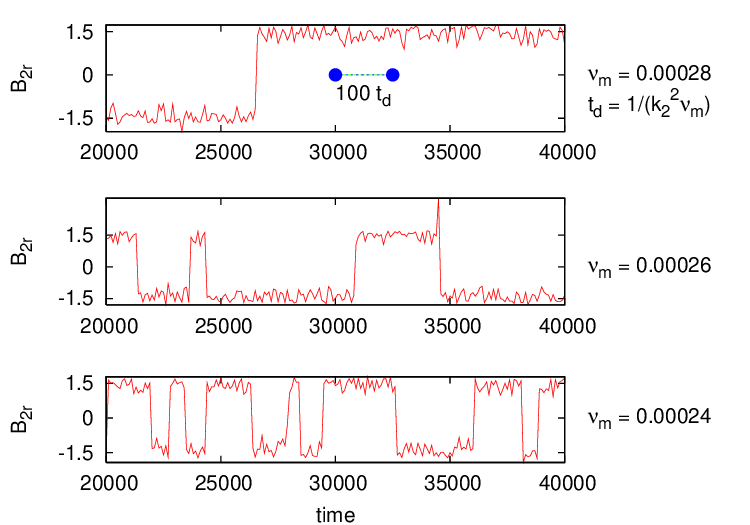}}
\caption{ 
Time behavior of $B_{2r}$ for three different values of $\nu_m$ (displayed on the left side) and constant $\nu$. The blue segment in the upper panel shows $100t_d$, where $t_d$ is the dissipative time scale computed
as $t_d=1./(k_2^2\nu_m)$. One time unit in the figure corresponds to the large scale eddy turnover time $1./(k_1|u_1|)$.
}
\label{fig2}
\end{figure}

\begin{figure}[h]
\begin{center}
\includegraphics[width=0.45\textwidth]{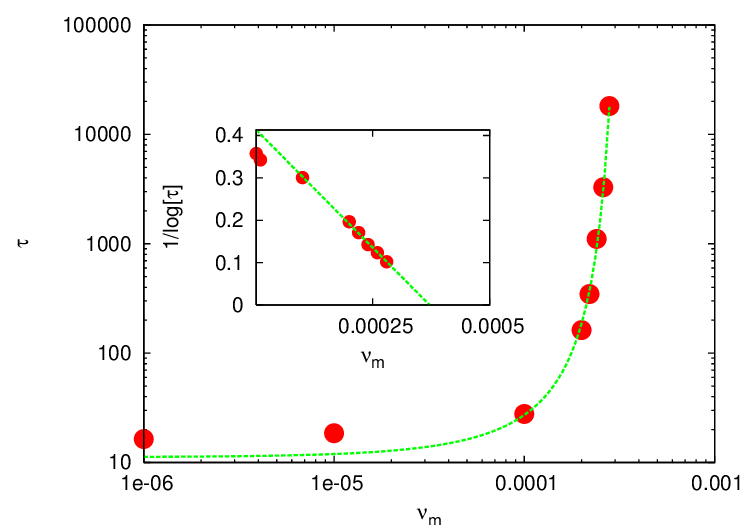}
\end{center}
\caption{Average persistence time $\tau$ as a function of the magnetic viscosity $\nu_m$ for $a_m=0.1$ and $\nu=10^{-7}$. The green line corresponds to the fit given by equation (\ref{fit}). In the insert we plot
$1./log(\tau)$ versus $\nu_m$ to highlight the linear behavior predicted by (\ref{fit}).
}
	\label{fig3}
\end{figure}

We are interested to study the behavior of the magnetic reversal, if any, as a function of $\nu_m$ and in particular in the region where $|B_2|$ saturates, i.e. it becomes independent of $\nu_m$. In figure~\ref{fig2}, we show three different time series of the $B_{2r} = Re(B_2)$ as a function of time for three different, relatively large, values of the magnetic diffusivity.  The figure highlights the two major informations discussed in this letter, namely the obervation of reversals between the two possible large scale equilibria and the dramatic increase of the time delay between reversals for increasing $\nu_m$  values. Note that this long time scale, as observed in the upper panel of figure~\ref{fig2}, is much longer than the characteristic time scale of $B_2$ near one of the two equilibrium states. The system spontaneously develops a significant time scale separation, for which given polarity is maintained for times much longer  than the magnetic diffusion time. In figure~\ref{fig3} we show the average reversal time as a function of $\nu_m$. More precisely, let us define $t_n$ the times at which $B_2(t_n)=0$ and $B_2$ has opposite sign before and after $t_n$. Then the reversal (or persisntance) is defined as $\tau_n \equiv t_n-t_{n-1}$, while the average reversal time $\tau$ is defined as the average of $\tau_n$. 

Figure~\ref{fig3} clearly shows that for large $\nu_m$, $\tau$ becomes extremely large (note that the figure is in log-log scale). Thus, even if neither $\langle |B_2|^2 \rangle$ nor $\epsilon_d$ depend on $\nu_m$, the effect of magnetic diffusivity is crucial for determining the average time reversal. In order to develop a theoretical framework aimed at understanding the result shown in figure~\ref{fig3},  we assume, in the region where $\langle |B_2|^2 \rangle$ is independent on $\nu_m$, that $B_{2i} \sim 0$ and that the term $F_2(u,B)$ can be divided  into an average forcing term proportional to $B_{2r}$ and a fluctuating part:
\begin{equation}
\label{approx}
F_2(u,B) = \beta B_2 + \phi'
\end{equation}
where $\beta$ depends on $f_0$ and $\phi'$ is supposed to be uncorrelated with the dynamics of $B_2$ , i.e. $ \langle [\phi' B_2^*] \rangle = 0$. Note that in the context of the mean-field approach to MHD, the first term $\beta B_2$ would correspond to an `alpha-effect'. Using (\ref{approx}), we can rewrite the equations for $B_2$ as follows:
\begin{equation}
\label{b2}
\frac{dB_2}{dt} = \beta B_2 - a_m B_2^3  + \phi' \ .
\end{equation}
where we neglect the dissipative term since $\beta \gg \nu_m k_2^2$ in the region of interest. Eq.(\ref{b2}) must be considered an {\it effective} equation describing the dynamics of the magnetic field $B_2$ and its reversals, and the fluctuations $\phi'$ incorporates the turbulent fluctuations from the velocity and magnetic field turbulent cascades. It is the effect of $\phi'$ which makes the system `jump' between the two statistically stationary states. Using (\ref{A}) we can obtain $\beta = \sqrt{A a_m}$ while the two statistical stationary states can be estimated as $\pm B_0$, $B_0^2 = \beta/a_m$. The effective equation (\ref{b2}) is a stochastically differential equation and, using large deviation theory, we can predict $\tau$ to be
\begin{equation}
\label{tau}
\tau  \sim \exp\left( \frac{\beta^2}{a_m \sigma} \right) = \exp\left( \frac{A}{\sigma} \right) \ ,
\end{equation}
where $\sigma$ is the variance of the noise $\phi'$ acting on the system. Let us notice that $A$ and $\sigma$ must have the same dimension, namely $[B]^2/{\rm time}$. Thus, we write $\sigma$ as  $\sigma= A f$ where $f$ is a function of the relevant dimensionless variables. In our problem the dimensionless numbers expected to play a role  for the dynamical behavior of the magnetic field are: the Reynolds number $R_V$, the magnetic Reynolds number $R_M$ (or equivalently the magnetic Prandtl number $P_M$) and the quantity $R_m = \sqrt{Aa_m/(\nu_m^2 k_2^4)}$ which is an effective Reynolds number, corresponding to the efficiency of energy transfers from the velocity field to the magnetic field at large scale. Given the fact that we operate at constant power input and $R_V= const$, we expect $f$ to be a function of $(R_m, R_M)$ only and we also expect the effective magnetic Reynolds number to be proportional to the integral one ($R_m \propto R_M$). We then show below that a very good description of our numerical results is obtained using the lowest order approximation $f(R_m, R_M) = R_M^* - R_M$, where $R_M^*$ is a critical magnetic Reynolds number below which reversals are not be observed. This choice leads to $\sigma = A (\nu_m^* - \nu_m)/uL$, and finally to
\begin{equation}
\label{fit}
\tau \sim \exp\left( \frac{C}{\nu_m^*-\nu_m} \right) \ ,
\end{equation}
where $C$ is a constant independent of $\nu_m$. This functional form is displayed in figure~\ref{fig3}; it agrees remarkably with the observed numerical values of $\tau$ for a rather large range. In the insert of figure~\ref{fig3} we show $1/log(\tau)$ as a function of $\nu_m$ to highlight the linear behavior predicted by aq.(\ref{fit}). The physical statement represented by (\ref{fit}) is that the average reversal time should show a critical slowing down for relatively large $\nu_m$.  In other words, we expect that fluctuations around the statistical equilibria  increase as $R_M$ increases. The increase of fluctuations may not be monotonic for very large $R_M$, which explains why we are not able to fit the entire range of $\nu_m$ shown in figure~\ref{fig3}. 
\begin{figure}
	\begin{center}
		\includegraphics[width=0.45\textwidth]{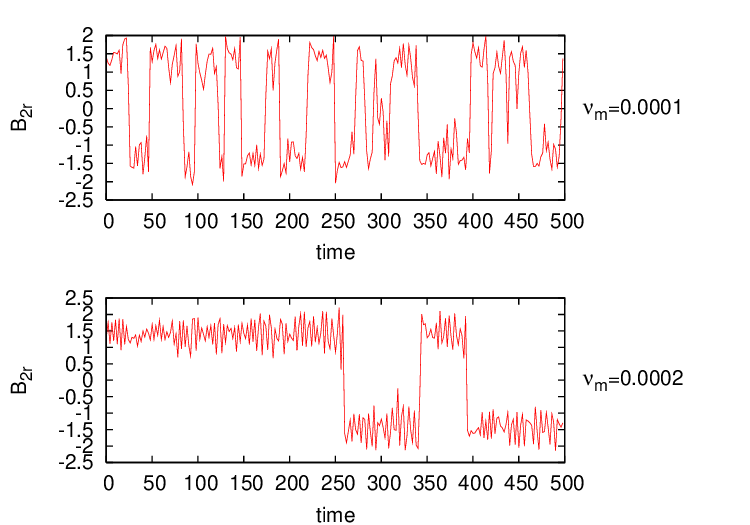}
	\end{center}
	\caption{
	Time behavior of $B_{2r}$ for two different values of $\nu_m$ (displayed on the right) obtained by using  a linear large scale equilibration $-\gamma B_2$ ($\gamma=0.13$) and by imposing $B_{2i}=0$.
	Note that although large scale equilibration is achieved by a linear damping on the magnetic field, $B_{2r}$ shows quite well defined statistical equilibria due to the symmetry breaking constrain $B_{2i}=0$.
	}
	\label{fig4}
\end{figure}

We finally comment on the choice of a non linear term in equation (4).   Actually, we can avoid non linear equilibration to obtain the same (qualitatively) results. In figure~\ref{fig4} we show two cases obtained with $M_2(B_2) = - \gamma B_2$ with the constrains $B_{2i}= 0$ and $\gamma = 0.13$. The equilibration mechanism is therefore linear while the symmetry breaking is obtained by the constrain $B_{2i}= 0$. Thus the two requirements, large scale dissipation and symmetry breaking, are satisfied. Figure~\ref{fig4} shows that statistical equilibria can be observed independent of non linear mechanism. Moreover, by changing the magnetic diffusivity,  we can still observe a rather large difference in the average reversal time. We argue that this effect is independent on the particular choice of  the equilibration mechanism since it is dictated  by dimensional analysis and large deviation theory.

{\bf Acknowledgments}  We thank Stephan Fauve for many interesting and critical discussions.


\end{document}